# Automated Thalamic Nuclei Segmentation Using Multi-Planar Cascaded Convolutional Neural Networks

Mohammad S. Majdi[1], Mahesh B. Keerthivasan[2,5], Brian K. Rutt[3], Natalie M. Zahr[4], Jeffrey J. Rodriguez[1], Manojkumar Saranathan[1,2]

**ABSTRACT**

**Purpose:** To develop a fast and accurate convolutional neural network based method for segmentation of thalamic nuclei.

**Methods:** A cascaded multi-planar scheme with a modified residual U-Net architecture was used to segment thalamic nuclei on conventional and white-matter-nulled (WMn) magnetization prepared rapid gradient echo (MPRAGE) data. A single network was optimized to work with images from healthy controls and patients with multiple sclerosis (MS) and essential tremor (ET), acquired at both 3T and 7T field strengths. Dice similarity coefficient and volume similarity index (VSI) were used to evaluate performance. Clinical utility was demonstrated by applying this method to study the effect of MS on thalamic nuclei atrophy.

**Results:** Segmentation of each thalamus into twelve nuclei was achieved in under a minute. For 7T WMn-MPRAGE, the proposed method outperforms current state-of-the-art on patients with ET with statistically significant improvements in Dice for five nuclei (increase in the range of 0.05-0.18) and VSI for four nuclei (increase in the range of 0.05-0.19), while performing comparably for healthy and MS subjects. Dice and VSI achieved using 7T WMn-MPRAGE data are comparable to those using 3T WMn-MPRAGE data. For conventional MPRAGE, the proposed method shows a statistically significant Dice improvement in the range of 0.14-0.63 over FreeSurfer for all nuclei and disease types. Effect of noise on network performance shows robustness to images with SNR as low as half the baseline SNR. Atrophy of four thalamic nuclei and whole thalamus was observed for MS patients compared to healthy control subjects, after controlling for the effect of parallel imaging, intracranial volume, gender, and age ($p<0.004$).

**Conclusion:** The proposed segmentation method is fast, accurate, performs well across disease types and field strengths, and shows great potential for improving our understanding of thalamic nuclei involvement in neurological diseases.

**KEYWORDS**

Convolutional neural network; Thalamic nuclei segmentation; Clinical analysis; White-matter-nulled MPRAGE

## 1. INTRODUCTION

The thalamus is a deep brain gray matter structure that relays information between various subcortical areas and the cerebral cortex [1] and plays a critical role in regulating sleep, consciousness, arousal, and awareness [2–4]. It is subdivided into multiple nuclei with varying functions. Thalamic involvement has been reported in schizophrenia [5,6], alcohol use disorder [7,8], Parkinson's disease [9], multiple sclerosis (MS) [10], and Alzheimer's disease [11]. These pathologies affect different thalamic nuclei differently and, therefore, accurate volumetry of thalamic nuclei can be beneficial for tracking disease progression and treatment efficacy [11,12]. An emerging application is in the treatment of essential tremor (ET) using deep brain stimulation [13–15]. Fast and accurate localization of the ventral intermediate (VIM) nucleus, whose abnormal electrical activity has been implicated in ET, can help improve the success rate of deep brain stimulation surgery and high-intensity focused ultrasound treatments of ET [16].

Manual delineation of thalamic nuclei from in-vivo scans is very tedious and requires specialized knowledge [17,18]. Due to low intra-thalamic contrast [19], thalamic nuclei are not easily distinguishable in conventional T1- and T2-weighted magnetic resonance images. As a result, most structural MRI based automated methods have only segmented the whole thalamus as part of subcortical brain segmentation [20–24]. Fischl et al. [20] used a voxel-wise probabilistic atlas of anatomy and MRI intensities to segment the brain into 15 subcortical structures, including the left and right thalamus (part of the FreeSurfer software [21]). Patenaude et al. [22] used active shape and appearance models along with a Bayesian framework (part of FSL package [25]).

Diffusion tensor imaging (DTI) based methods that use either local or global properties of the diffusion tensor have been more popular for thalamic nuclei segmentation. Behrens et al. [26] used tractography of cortical projections to the thalamus to segment thalamic regions, but this method requires precise knowledge of neuroanatomy to identify the relevant cortical regions. More automated, computationally efficient methods have been proposed that use k-means clustering of the dominant diffusion orientation to achieve thalamic parcellation [27–29]. The most consistent DTI method [30] to date uses spherical harmonic decomposition based orientation distribution functions to achieve robust segmentation of seven thalamic nuclei. However, the low spatial resolution of echo-planar imaging which underlies DTI and the predominance of gray matter in the thalamus which results in low anisotropy make these DTI-based methods suboptimal [31], often resulting in segmentation of only the larger thalamic groups. Advanced techniques such as susceptibility-weighted imaging (SWI) [32] can provide better intra-thalamic contrast and have been used for segmentation of thalamic nuclei at 7T [33,34]. However, these methods have found limited application in presurgical targeting, focusing mainly on the VIM nucleus.

Other attempts to segment thalamic nuclei have relied on supervised machine learning techniques. Stough et al. [35] used a random forest classifier to combine features extracted from DTI along with traditional T1-weighted MRI to segment the four large thalamic groups along with the lateral geniculate nucleus (LGN) and medial geniculate nucleus (MGN). Glaister et al. [36] extended the above approach by proposing a cascaded random forest classifier that makes use of DTI features along with atlas-based nuclei priors to first detect the thalamus followed by nuclei segmentation.

Recently, high spatial resolution structural MRI has been investigated for thalamic nuclei segmentation. The most widely used T1-weighted structural MRI sequence is magnetization prepared rapid gradient echo (MPRAGE), where the cerebrospinal fluid (CSF) is nulled. We refer to this method as CSFn-MPRAGE. Iglesias et al. [24] proposed a probabilistic atlas constructed using manual delineation of 26 thalamic nuclei per thalamus on six autopsy specimens and used Bayesian inference to segment 3T MPRAGE images into 26 nuclei per side [37,38]. However, this method is very time consuming, requiring multiple hours for the segmentation of one subject and has not been thoroughly validated against manual segmentation [24]. Liu et al. [36,39] segmented thalamic nuclei from 3T T1-weighted MRI data using an atlas developed from multiple MPRAGE and SWI sequences acquired at 7T. A multi-atlas label fusion and statistical shape modeling algorithm was used to transfer from 7T to 3T. Variants of the MPRAGE sequence have been proposed to better visualize the intra-thalamic structures [40,41]. Su et al. [31] used a WMn-MPRAGE sequence that is optimized for intra-thalamic contrast [19] in conjunction with a multi-atlas technique, called thalamus optimized multi-atlas segmentation (THOMAS), to segment the thalamus into 12 nuclei. The performance of this based method hinges on the accuracy of a computationally expensive registration step [24,42]. This method has only been validated on specialized WMn-MPRAGE data.

In recent years, deep learning has become a popular tool for detection and segmentation of medical images [43–48]. Convolutional neural networks (CNNs) are a class of deep learning techniques that use convolutional kernels to capture the non-linear mapping between an input image and its segmentation labels. Unlike atlas-based segmentation techniques, CNNs do not depend on image registration and manual feature extraction [49]. While many studies have explored the advantages of using CNNs for subcortical segmentation [43,46,50,51], those studies are limited to the whole thalamus. Due to a paucity of training data and the high computational and memory requirements of 3D analysis, most proposed methods make use of 2D CNNs [50]. However, the use of 2D networks does not fully exploit the anatomical information present in 3D MRI data. Alternatively, multi-planar techniques that make use of 2D CNNs along the three orthogonal planes have been shown [46,52–54] to improve segmentation performance with lower computational cost than a full 3D analysis. Transfer learning techniques have also been investigated to mitigate the lack of sufficient training data [55].

In this work, we propose the use of a modified residual U-Net in a cascaded multi-planar scheme for thalamic nuclei segmentation. We first demonstrate this method for WMn-MPRAGE [19] data, evaluating it on data from 7T as well as 3T. We then extend this work to the more commonly acquired CSFn-MPRAGE by fine tuning the network trained on WMn-MPRAGE data. The performance of both networks is validated on healthy subjects and patients with MS and ET and compared to current state-of-the-art segmentation methods. Finally, robustness of the proposed method to SNR and its applicability to data from patients with multiple sclerosis are investigated.

## 2. METHODS

### 2.1 Network Architecture

The proposed method uses a modified residual U-Net architecture (mRU-Net) for 2D thalamic nuclei segmentation as shown in Figure 1 (top panel). The conventional U-Net [56] uses a contracting (encoder) layer and an expanding (decoder) layer for multi-resolution feature extraction and synthesis. To improve the convergence performance and reduce overfitting, we incorporated batch normalization [57] and dropout [58] layers into the network. Residual convolutional blocks [59] were included to mitigate the problem of vanishing and exploding loss function gradients. Each convolutional block consists of 2D convolution units followed by a batch normalization and a leaky rectified linear unit [60] (leakage factor 0.1). A 1x1 convolution with a sigmoid activation function maps the final feature maps into the desired number of classes, generating a probability map for each class.

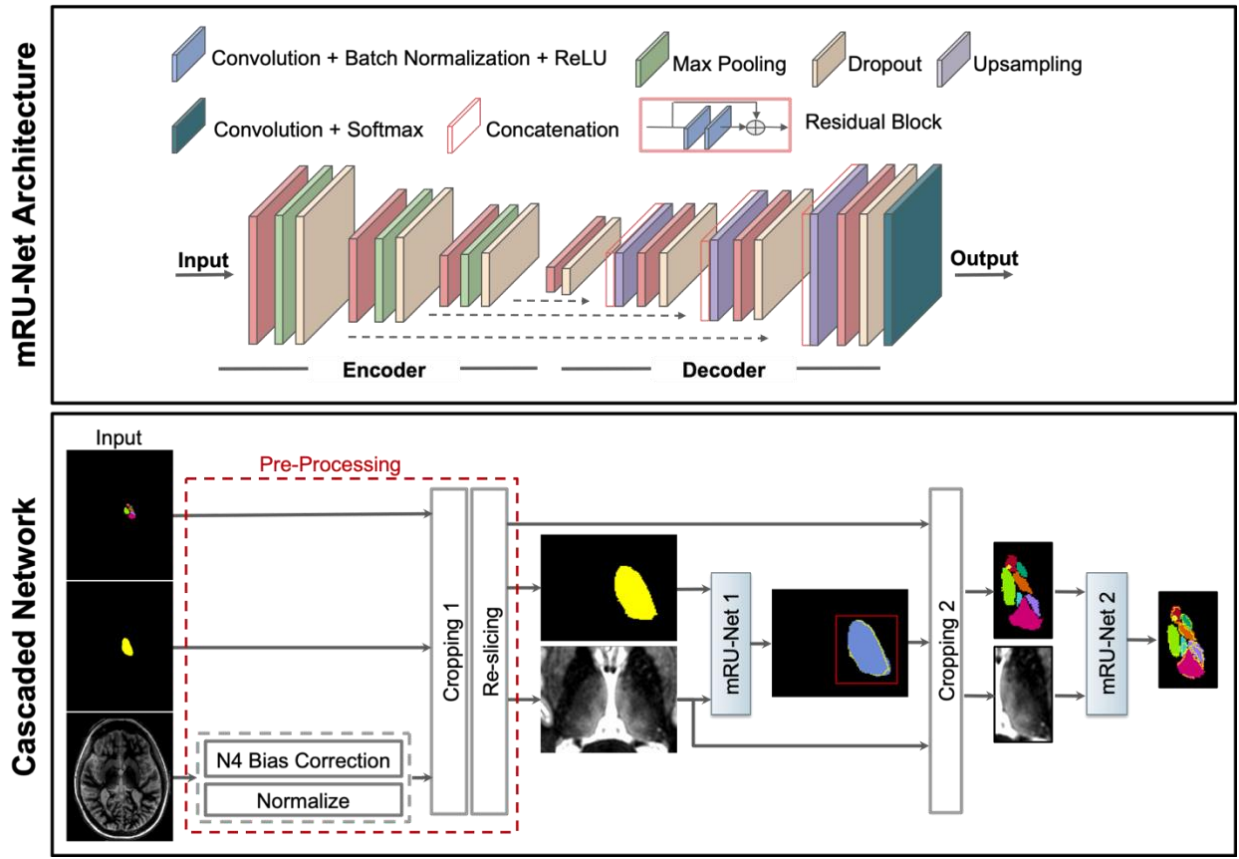

Figure 1: Top panel shows the mRU-Net architecture, where normalization and residual blocks are added to a standard U-Net. Bottom panel shows the proposed cascaded network for thalamic nuclei segmentation. Following pre-processing, mRU-Net1 is trained to segment

For thalamic segmentation, we use a two-step cascaded approach [44], as shown in Figure 1 (bottom panel). A first mRU-Net initially segments the whole thalamus, and this guides a second mRU-Net, which segments the different thalamic nuclei. Pre-processed 2D images (see section 2.3 for details) along with their manual segmentation masks are used to train the first network to segment the whole thalamus. A bounding box encompassing the whole thalamus is used to crop the input image (and its corresponding manual segmentation masks) before it is input to the second network to perform thalamic nuclei segmentation. Note that an optional pre-processing cropping step ("Cropping 1" in Figure 1) is also added prior to inputting images to the first network.

Due to the limitation in the number of labeled datasets available, training a generalizable 3D CNN which can exploit the 3D structure is infeasible. In order to overcome this limitation, and take advantage of the spatial information present in an isotropic-resolution 3D MRI dataset, we use a multi-planar [54] approach as shown in Figure 2. The input 3D dataset is pre-processed, reformatted into three orthogonal orientations (axial, coronal, and sagittal), and the resulting 2D images are fed into three cascaded 2D networks (from Figure 1) to take advantage of the complementary information from each orientation. For each cascaded network, the output segmentations are reformatted to the original imaging orientation and then fused using voxel-wise majority voting.

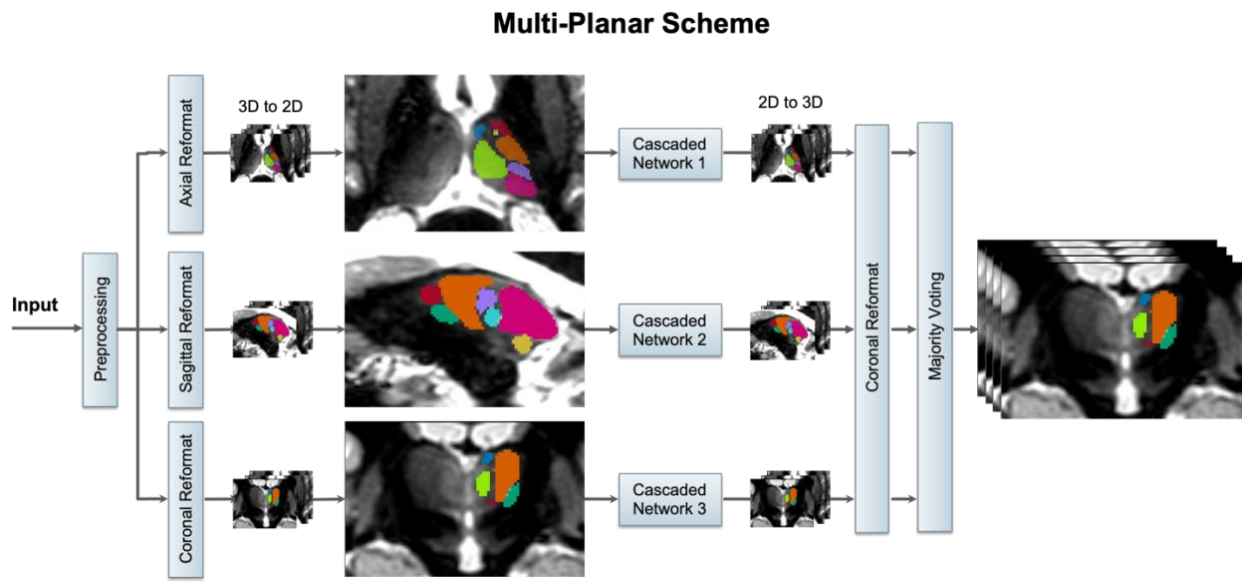

Figure 2: Schematic of the multi-planar scheme, based on the cascaded network of Figure 1. The 3D input image is reformatted to sagittal, coronal, and axial planes and fed as 2D inputs to three separate cascaded networks. The segmentations from the three 2D networks are fused using majority voting to generate the final 3D thalamic nuclei segmentation.

In the bottom panel of Figure 1, the first network (whole thalamus segmentation) uses a Dice [61] loss function along with sigmoid activation. For the second network, we investigated two different loss functions with a one-hot encoding approach in treating multiple classes: a) Weighted cross-entropy and b) Dice.

Dice loss function is defined as follows:

$$\text{Loss} = \frac{1}{C}\sum_{i=0}^{C-1}(1 - \text{Dice}(A_i, B_i)) \tag{1}$$

$$\text{Dice}(A_i, B_i) = \frac{2|A_i \cap B_i|}{|A_i \cup B_i|} \tag{2}$$

where $A_i$ and $B_i$ are the automated and manual segmentations for nucleus $i$ and $|.|$ is the cardinality of a set (i.e., number of pixels). Note that the different classes are implicitly weighted when using a Dice-based loss function [62,63]. A weighted cross entropy was used to mitigate the effect of class imbalance, where the weight for each class is defined as the ratio of the number of foreground to background voxels in the manual segmentation mask. The weighted cross-entropy loss is defined as

$$\text{Loss} = -\sum_{i=0}^{C-1} w_i ( q_i \log(p_i) + (1 - q_i) \log(1 - p_i)) \tag{3}$$

where $C$ is the number of classes including the background (2 for the whole-thalamus network or 13 for the subsequent thalamic nuclei segmentation network for 12 nuclei), $p_i = \exp(\hat{y}_i)/\sum_{k=0}^{C-1}\exp(\hat{y}_k)$ is the posterior probability for class $i$ obtained by applying the sigmoid function to the network's final feature maps $\hat{y}_i$, $w_i$ represents the weight for class $i$, and $q_i$ is the ground truth for class $i$.

Two instances of the cascaded multi-planar network were implemented: a WMn network for segmenting WMn-MPRAGE images, and a CSFn network for segmenting CSFn-MPRAGE images.

### 2.2 Datasets and Labels

To evaluate thalamic nuclei segmentation of WMn-MPRAGE images, we used data from 40 subjects (13 healthy subjects, 15 patients with MS, and 12 patients with ET) acquired on a 7T scanner. We used data from the 12 ET patients rescanned on a 3T scanner to extend this work to 3T. To evaluate thalamic nuclei segmentation of CSFn-MPRAGE images, we used data from 33 subjects (6 healthy subjects, 15 patients with MS, and 12 patients with ET) scanned with both WMn-MPRAGE and CSFn-MPRAGE pulse sequences on a 7T scanner. A separate dataset consisting of 93 WMn-MPRAGE images from a 3T scanner was used just for the purpose of network initialization (referred to as initialization dataset); this included patients with alcohol use disorder and healthy control subjects. All subjects were scanned on GE (General Electric Healthcare, Waukesha WI) 7T and 3T scanners after obtaining prior informed consent and adhering to institutional review board protocols. The pulse sequence parameters for both the WMn-MPRAGE and CSFn-MPRAGE pulse sequences are listed in Supporting Table S1.

**Supporting Table 1:** Image acquisition parameters for WMn- and CSFn-MPRAGE for 7T and 3T data used in this study.

|  | WMn-MPRAGE | | CSFn-MPRAGE |
|---|---|---|---|
| **Scanner** | 3T (GE) | 7T (GE) | 7T (GE) |
| **TR/TS (ms)** | 10/4500 | 10/6000 | 7.2/3000 |
| **TI (ms)** | 500 | 680 | 1200 |
| **Flip (deg)** | 9 | 4 | 6 |
| **Matrix** | 180 × 220 × 180 | 180 × 220 × 180 | 180 × 220 × 180 |
| **Acq. resolution (mm)** | 1x1x1 | 1x1x1 | 1x1x1 |
| **Rec. resolution (mm)** | 0.7x0.5x0.7 | 0.7x0.5x0.7 | 0.7x0.5x0.7 |
| **Parallel imaging** | None | 1.5 × 1.5 | 3x1 |
| **Coil** | 8-channel GE | 32-channel Nova | 32-channel Nova |

Reference labels for WMn-MPRAGE images were generated using manual segmentation of thalamic nuclei, performed by a trained neuroradiologist using the Morel histological atlas [64] as a guide. Eleven thalamic nuclei, the whole thalamus, and the mammillothalamic tract (MTT) were delineated using freehand spline drawing tools to build the 3D vector-based model of each structure.

The eleven delineated nuclei are grouped as follows:
  (i) **medial group**: mediodorsal (MD), centromedian (CM), habenula (Hb)
  (ii) **posterior group**: pulvinar (Pul), medial geniculate nucleus (MGN), lateral geniculate nucleus (LGN)
  (iii) **lateral group**: ventral posterolateral (VPL), ventral lateral anterior (VLa), ventral lateral posterior (VLp), ventral anterior nucleus (VA)
  (iv) **anterior group**: anteroventral (AV)

Reference labels for CSFn-MPRAGE images were obtained by affine registering WMn-MPRAGE images to the corresponding CSFn-MPRAGE images for each subject and warping the labels using nearest-neighbor interpolation. Reference labels for the initialization dataset were obtained using THOMAS segmentation [31]. Due to the laborious nature of manual segmentation, only left thalamic nuclei were manually segmented. As a result, all networks were trained only on left thalamic nuclei. To delineate the right thalamic nuclei, the input images were flipped in the left-right direction and segmented using the networks trained on the left manual segmentation labels and then flipped back.

### 2.3 Network Training

All networks were implemented in Python and Keras [65] with a TensorFlow backend using an NVIDIA P100 GPU with 16 GB GDDR5 RAM (**source code** available at https://github.com/artinmajdi/Thalamic-Nuclei-Segmentation). The WMn network was trained using combined 3T and 7T data. The CSFn network was initialized using this WMn network, and then fine-tuned (i.e. re-trained using WMn labels registered to CSFn data) to adapt to CSFn-MPRAGE contrast. The networks were trained using Adam optimizer [66] with 300 epochs and a batch size of 100. Network performance was evaluated for both weighted cross entropy and Dice loss functions. The number of layers, number of convolutional feature maps, and learning rates were chosen based on hyper-parameter tuning. Further, a scheduler was used to set the learning rate in each epoch, starting from an initial value of 0.001 that reduces by a factor of 0.5 if the validation Dice plateaus for 15 epochs. In the multi-planar approach, after a series of hyperparameter tuning, the number of feature maps in the first layer was set to 40, 30, and 20 for the sagittal, coronal and axial networks, respectively. The number of feature maps in each succeeding layer was increased by a factor of two as proposed in the conventional U-Net [56]. For the WMn and CSFn networks, 20% and 25% of data were randomly selected for validation, while the remaining 80% and 75% were used for training, respectively.

### 2.4 Pre- and Post-Processing

All input images were pre-processed by performing N4 bias-field correction and zero mean unit standard deviation normalization. They were then reformatted to a standard imaging plane and resolution (axial with 0.7 × 0.5 × 0.7 mm3). To ensure enough variability, the training data were augmented using random in-plane rotations of up to ±7° in three different planes, creating six new images in each orientation and a net increase in the number of training data by a factor of 18. To reduce the memory and computation power requirements, a cropping step was added to the pre-processing as an extra option. To automate this cropping step, we employed a WMn-MPRAGE template created from mutual registration and averaging of 20 prior WMn-MPRAGE datasets as described in Su et al. [31]. A bounding box covering both thalami was manually drawn once, on this template and warped into the input image space by affine registering the template with the input image. The pre-processing steps are shown in Figure 1 (bottom panel). Typical input sizes for the original, pre-processed (input to the first whole thalamus network), and cropped images (input to the second nuclei segmentation network) were 256 ×373 × 256, 97 × 94 × 63, and 52 × 84 × 48, respectively.

### 2.5 Performance Evaluation Measures

The automated segmentation performance was evaluated with respect to the manual delineations using Dice similarity coefficient [67], and volume similarity index (VSI) [31]. The Dice similarity coefficient (shortened as Dice) shown in Eq. (2) measures overlap between the manual segmentation and the automated segmentation computed by the network. The volume similarity index is defined as

$$\text{VSI}(A, B) = \frac{1}{C} \sum_{i=0}^{C-1} \left(1 - \frac{||A_i| - |B_i||}{|A_i| + |B_i|}\right) \quad (4)$$

where $A_i$ and $B_i$ are the automated and manual segmentations for nucleus $i$, with a VSI of 1 indicating identical volumes.

The outputs from the network were binarized using thresholds computed from precision-recall curves. The threshold (0.7) was determined by finding the tradeoff between values of precision and recall that would correspond to the lowest false positive and false negative rates.

## 2.6 Experiments

### 2.6.1 Network Optimization

To potentially reduce memory and computational burden, we explored an additional step to crop the input images of the first thalamus segmentation network ("Cropping 1" in the lower panel of Figure 1). The performance of the cascaded scheme was also compared to a non-cascaded scheme which is a network similar to the proposed cascaded scheme with the exception of *removing* the second network's cropping step ("Cropping 2" in Figure 1)). Finally, the effect of network initialization was studied by training two separate networks, one using random initialization and the other initialized using weights from a separate network that was trained on 3T WMn-MPRAGE data with THOMAS reference labels (3T initialization network). Lastly, to find the optimal loss function, two networks were trained separately using weighted binary cross entropy and Dice loss functions, respectively. All the above experiments were performed on a subset (n=11) of subjects randomly chosen from the WMn-MPRAGE dataset and equitably distributed across control subjects and disease types, except for cropping experiments which were done on the entire dataset (n=40).

### 2.6.2 Network Performance

To evaluate the segmentation performance of the optimized multi-planar cascaded networks, a 4-fold and 8-fold cross validation was performed using approximately 20% (out of 52 cases) and 25% (out of 33 cases) for each fold on WMn-MPRAGE and CSFn-MPRAGE data, respectively. Data in cross validation folds were equitably distributed with respect to control/disease type. To extend the applicability of the CNN based method to 3T, data from the 12 ET patients rescanned on a 3T scanner were used as part of the cross validation along with all of the 7T WMn-MPRAGE data. The segmentation results of our method were compared to THOMAS and FreeSurfer based segmentations for WMn-MPRAGE and CSFn-MPRAGE images, respectively using Dice and VSI measures.

## 2.7 Network Robustness to Noise

Many factors influence the MR image signal-to-noise ratio (SNR) including the receiver coil resistance, inductive losses in the sample [68], image voxel size, receiver bandwidth [69], and pulse sequence parameters. To study the robustness of our method to different levels of SNR, randomly generated noise was incorporated using Eq (5) into a WMn-MPRAGE magnitude image $I$ (resulting in a Rician noise corrupted image $I'$).

$$I' = \sqrt{(I + n_{\text{real}})^2 + n_{\text{imag}}^2} \qquad (5)$$

where $n_{\text{real}}$ and $n_{\text{imag}}$ are independent Gaussian distributed random variables $N(0, \sigma^2)$,. Thalamic SNR is measured as the ratio of the mean thalamic signal to standard deviation of the noise from two region of interests (ROIs) placed over the thalamus and the image background. Starting from the original acquired WMn-MPRAGE images (nominal thalamic SNR of 23.5), noise with increasing standard deviation $\sigma$ was added to produce 10 images with an SNR in the range of 23.5-8 (i.e. ~ 3X degradation in thalamic SNR).

### 2.8 Clinical Analysis in Multiple Sclerosis

To assess the effect of multiple sclerosis on specific thalamic nuclei, analysis of covariance (ANCOVA) was performed to compare thalamic nuclei volumes between healthy controls and patients with MS, controlling for age, gender, parallel imaging (PI), and intracranial cavity volume (ICV). The estimated total intracranial volume (eTIV) from FreeSurfer segmentation was used for ICV to account for differing head sizes. Left and right thalamic nuclei were segmented from 7T WMn-MPRAGE data acquired on 15 MS patients (13 patients had relapsing-remitting MS, while 2 patients had secondary-progressive MS) and 13 healthy subjects (free of neurologic, psychiatric, or systemic diseases, and drug or alcohol abuse). A Bonferroni corrected p-value of 0.05/11=0.005 (to account for multiple comparisons of the 11 nuclei) was used.

## 3. RESULTS

### 3.1 Network Optimization

**Training Time**

The training time required on one NVIDIA P100 GPU card for the whole thalamus and multi-class thalamic nuclei segmentation was 1 hour and 1.5 hours, respectively, for a single imaging orientation. The cumulative training time for the multi-planar cascaded scheme was (1 + 1.5) × 3 = 7.5 hours (accounting for axial, coronal, and sagittal orientations). This number can be reduced to 2.5 hours if the training in each orientation is performed in parallel (three GPUs, one for each orientation). The time required for pre-processing and segmentation of each subject in the testing phase of the final multi-planar scheme was 3 min. and 1 min., respectively. The use of the cascaded scheme reduced the required memory for training by up to 86%, enabling the use of augmented data during the training process. Even though the cascaded scheme was trained in the presence of 15 × augmented data, the number of epochs and overall time of convergence during training was reduced by 66% and 21%, respectively, in comparison to the non-cascaded algorithm.

**Cropping, Initialization, Loss Function**

For the 40 test subjects (13 control, 15 MS, 12 ET subjects), no statistical difference in average Dice was observed between using uncropped and cropped input images to the thalamic segmentation network (Figure 1) while a 93% reduction of memory requirements was achieved using the initial cropping step ("Cropping 1" in Figure 1). Following this experiment, all remaining experiments included this initial cropping step as part of pre-processing.

The WMn network initialized using weights from a 3T initialization network showed a significant improvement in Dice for two nuclei (VA, and Hb) and an increase in convergence rate compared to random initialization (Supporting Figure S1). Following this, all experiments on WMn-MPRAGE data were performed using this initializing step. For the WMn network, the use of the Dice loss function showed a statistically significant improvement in Dice for the whole thalamus and 4 nuclei (VPI, LGN, CM, MGN) compared to a weighted cross entropy loss function (Supporting Table S2). Further, it reduced the overall number of epochs, required training time (per epoch) and convergence time by 32%, 16%, and 43%, respectively. As a result, the Dice loss function was used for all further experiments.

**Supporting Table 2:** Effect of loss function on Dice and VSI. Average Dice and VSI shown for 11 subjects.

| | Dice | | VSI | |
|---|---|---|---|---|
| **Nuclei** | Dice Loss | BCE Loss | Dice Loss | BCE Loss |
| **Thalamus** | 0.92† | 0.91 | 0.98† | 0.96 |
| **Pul** | 0.86 | 0.85 | 0.97 | 0.96 |
| **VLP** | 0.78 | 0.79 | 0.95 | 0.95 |
| **MD-Pf** | 0.85 | 0.85 | 0.95 | 0.94 |
| **VPI** | 0.64† | 0.61 | 0.92 | 0.89 |
| **VA** | 0.70 | 0.69 | 0.92 | 0.93 |
| **AV** | 0.76 | 0.74 | 0.87 | 0.85 |
| **VLa** | 0.69 | 0.67 | 0.92 | 0.89 |
| **CM** | 0.70† | 0.65 | 0.93† | 0.86 |
| **LGN** | 0.70† | 0.66 | 0.90 | 0.85 |
| **MGN** | 0.70† | 0.64 | 0.89 | 0.81 |
| **MTT** | 0.68 | 0.63 | 0.88 | 0.81 |
| **Hb** | 0.77 | 0.75 | 0.89 | 0.88 |

† $p < 0.05$ BCE vs. Dice loss function

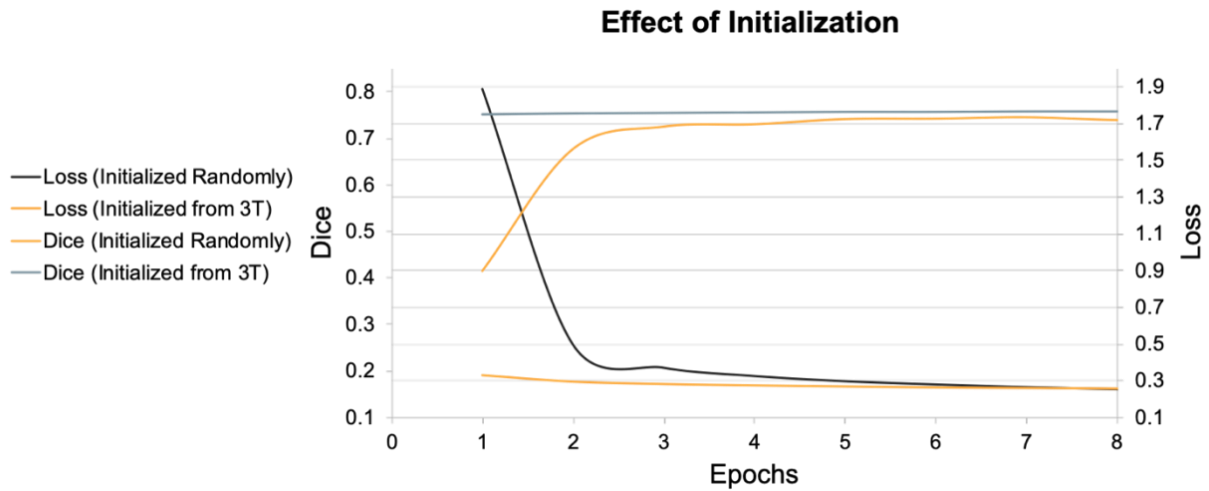

**Supporting Figure 1:** Effect of initialization on final accuracy and convergence curves. validation accuracy (Dice) and training loss for a network with random initialization vs. initialization from a network trained on a 3T dataset is shown. The segmentation results indicate a statistically significant improvement over two nuclei (VA and Hb) when initialized from the 3T network.

**3.2 Network Performance**

Figure 3 compares shows Dice for (a) WMn-MPRAGE data segmented using the proposed method (blue) and THOMAS (red) and (b) CSFn-MPRAGE data segmented using the proposed method (blue) and FreeSurfer (red). Note that this aggregates the Dice over control, MS, and ET cases (WMn: n=40 and CSFn: n=33). Our proposed CNN-based method shows improved Dice over THOMAS and FreeSurfer for seven and ten nuclei, respectively.

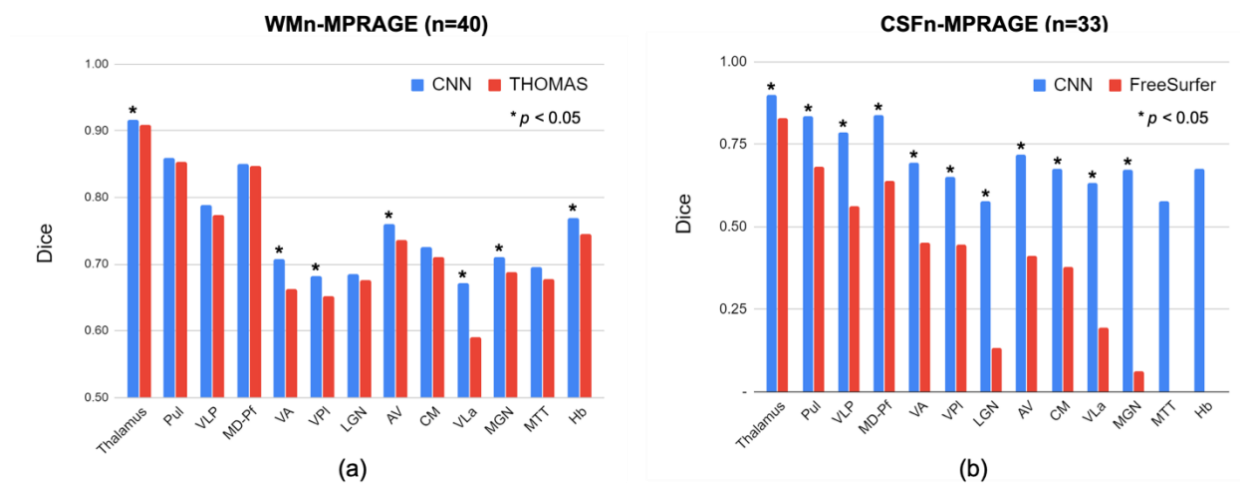

**Figure 3:** Comparison of average Dice for a) WMn-MPRAGE and b) CSFn-MPRAGE for the proposed method (blue) and THOMAS/FreeSurfer (red), aggregated over all cases. Dice values segregated by disease type is shown in Tables 1 and 2.

**WMn-MPRAGE Data**

Table 1 reports average Dice and VSI for the proposed method against THOMAS on WMn-MPRAGE data separately for control subjects (n=13), MS patients (n=15) and ET patients (n=12). Note that the thalamic nuclei are arranged in descending order of size with the smaller nuclei (<200 mm$_3$) shaded in gray. This format is used for all subsequent tables. Average Dice for the ET subjects showed significant improvements for the whole thalamus and 5 nuclei compared to THOMAS, ranging from 0.02 (whole thalamus) to 0.17 (VLa). Average VSI showed improvements for 4 nuclei, ranging from 0.05 (VLp) to 0.18 (VLa). For the control and MS subjects, the Dice and VSI were largely comparable, with a modest but statistically significant improvement in average Dice over 5 nuclei (range 0.03-0.06) and 3 nuclei (0.01-0.04) for control and MS subjects, respectively. Average VSI showed improvements for 2 nuclei (0.11) while performing worse for 1 nucleus (0.03) for control subjects and no differences for MS patients.

Average Dice and VSI for the proposed method on 3T WMn-MPRAGE data from ET patients (same patients that were scanned at 7T) showed no statistically significant differences in Dice or VSI between the 7T and 3T data, attesting to the ability of the network to be useful in clinically relevant field strengths.

Figure 4 shows segmentation results from a patient with ET scanned on a 7T (top panel) and a 3T MRI scanner (bottom panel) using WMn-MPRAGE. Representative axial and coronal slices with automated segmentation labels overlaid for both the left and right thalamus are shown on the right. The increased SNR and $B_1$ inhomogeneity in the 7T image (white arrow) can be clearly seen at the periphery while the inter-nuclear contrast seems comparable. The Dices (and hence the segmentations) are virtually identical as shown in Table 1.

**CSFn-MPRAGE Data**

Table 2 shows the average Dice and VSI for the proposed method and FreeSurfer on CSFn-MPRAGE data separately for 6 healthy control subjects, 15 patients with MS, and 12 patients with ET. The proposed method significantly outperformed FreeSurfer on all nuclei and disease types with an average improvement in Dice of 0.33, 0.26, and 0.30 for control, MS, and ET subjects, respectively. VSI showed improvements over 3, 6, and 4 nuclei with an average improvement of 0.39, 0.37, and 0.23 over control, MS, and ET subjects, respectively.

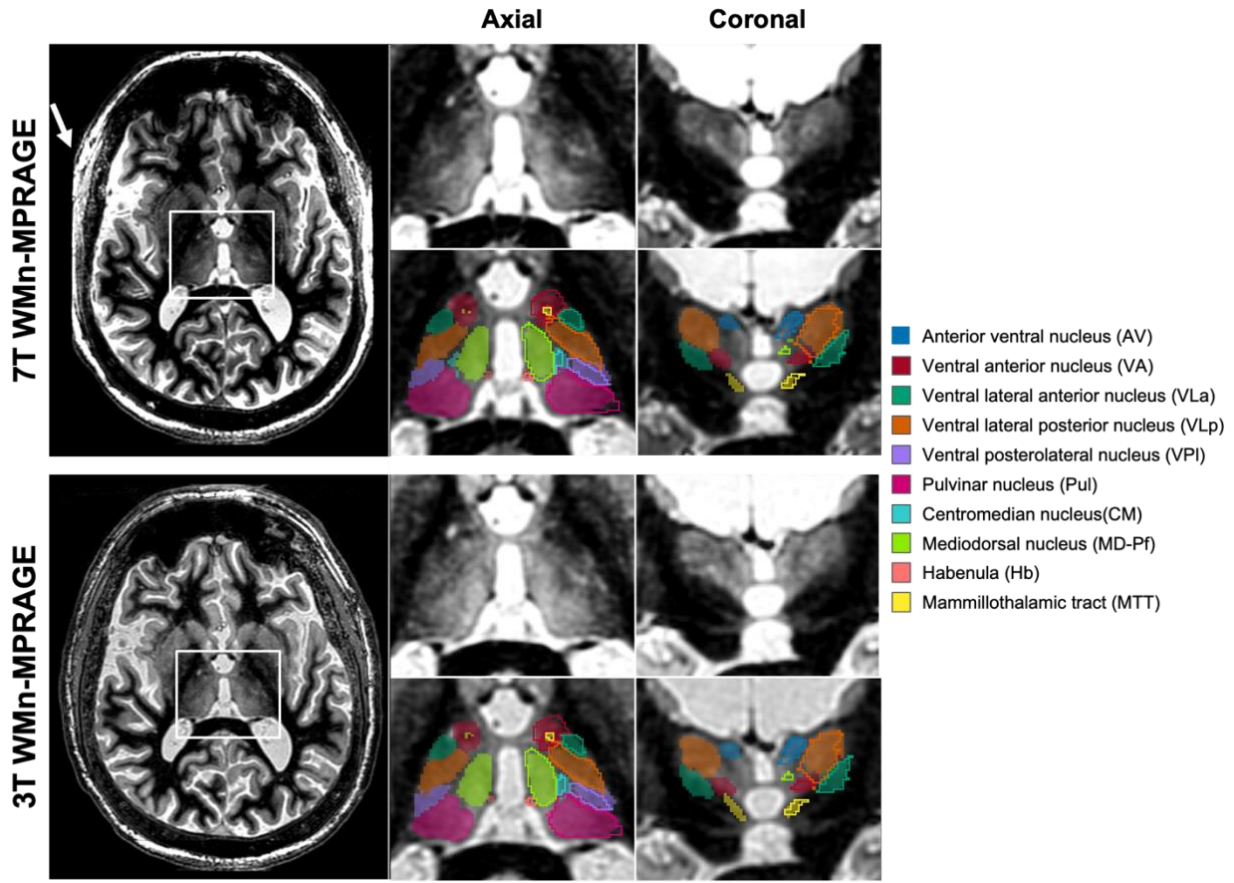

**Figure 4:** WMn-MPRAGE images from a patient with ET acquired at 7T (top left) and 3T (bottom left). The zoomed inset images on the right show representative axial and coronal sections with and without automated segmentation overlays. The automated thalamic nuclei segmentations are almost identical to the manual segmentation shown for the left thalamus in outlines. Note the clear visualization of small structures like the MTT and habenula on both sets of images. The increased B1 heterogeneity at 7T can be clearly seen (white arrow) but the cropped images are comparable.

**Table 1:** Comparison of mean Dice and VSI for THOMAS vs. proposed method for WMn-MPRAGE data (median volumes in mm3 are shown in parentheses and nuclei with <200 mm3 shaded, red bold indicates >10% change)

| | Dice | | | | | | | VSI | | | | | | |
|---|---|---|---|---|---|---|---|---|---|---|---|---|---|---|
| Disease | Control | | MS | | ET-7T | | ET-3T | Control | | MS | | ET-7T | | ET-3T |
| Nuclei (Volume) | THOMAS (n=13) | CNN (n=13) | THOMAS (n=15) | CNN (n=15) | THOMAS (n=12) | CNN (n=12) | CNN (n=12) | THOMAS (n=13) | CNN (n=13) | THOMAS (n=15) | CNN (n=15) | THOMAS (n=12) | CNN (n=12) | CNN (n=12) |
| Thalamus (5515) | 0.92 | 0.92 | 0.92 | 0.92 | 0.88 | 0.90* | 0.90 | 0.97 | 0.97 | 0.98 | 0.98 | 0.94 | 0.95 | 0.96 |
| Pul (1286) | 0.87 | 0.87 | 0.85 | 0.86* | 0.84 | 0.85 | 0.84 | 0.96 | 0.96 | 0.95 | 0.95 | 0.95 | 0.95 | 0.95 |
| VLp (882) | 0.79 | 0.79 | 0.79 | 0.78 | 0.75 | 0.80* | 0.79 | 0.97* | 0.94 | 0.93 | 0.93 | 0.89 | 0.94* | 0.95 |
| MD-Pf (686) | 0.86 | 0.85 | 0.85 | 0.86 | 0.83 | 0.84 | 0.84 | 0.94 | 0.92 | 0.95 | 0.95 | 0.93 | 0.93 | 0.94 |
| VPl (346) | 0.68 | 0.71* | 0.69 | 0.69 | 0.57 | **0.64*** | 0.63 | 0.90 | 0.92 | 0.92 | 0.91 | 0.86 | 0.92 | 0.92 |
| VA (301) | 0.71 | 0.74* | 0.69 | 0.72* | 0.58 | **0.66*** | 0.63 | 0.88 | 0.89 | 0.91 | 0.88 | 0.92 | 0.95 | 0.93 |
| AV (146) | 0.77 | 0.79 | 0.78 | 0.77 | 0.64 | **0.71*** | 0.71 | 0.89 | 0.93 | 0.93 | 0.88 | 0.71 | **0.84*** | 0.92 |
| VLa (121) | 0.65 | 0.71* | 0.64 | 0.66 | 0.47 | **0.64*** | 0.63 | 0.81 | **0.92*** | 0.87 | 0.86 | 0.69 | **0.87*** | 0.92 |
| CM (119) | 0.75 | 0.75 | 0.74 | 0.75 | 0.63 | 0.66 | 0.66 | 0.91 | 0.94 | 0.89 | 0.89 | 0.92 | 0.91 | 0.92 |
| LGN (115) | 0.75 | 0.76 | 0.71 | 0.69 | 0.55 | 0.59 | 0.59 | 0.91 | 0.92 | 0.91 | 0.89 | 0.82 | 0.88* | 0.90 |
| MGN (76) | 0.68 | 0.74* | 0.75 | 0.74 | 0.63 | 0.65 | 0.65 | 0.79 | **0.90*** | 0.90 | 0.87 | 0.77 | 0.83 | 0.86 |
| MTT (49) | 0.67 | 0.71 | 0.69 | 0.71 | 0.67 | 0.66 | 0.64 | 0.91 | 0.88 | 0.91 | 0.89 | 0.89 | 0.87 | 0.89 |
| Hb (29) | 0.76 | 0.80* | 0.72 | 0.76* | 0.76 | 0.74 | 0.72 | 0.88 | 0.92 | 0.90 | 0.91 | 0.89 | 0.88 | 0.87 |

\* p < 0.05 THOMAS vs. CNN

**Table 2:** Comparison of mean Dice and VSI for FreeSurfer vs. proposed method for CSFn-MPRAGE data (median volumes in mm3 are shown in parentheses and nuclei with <200 mm3 shaded, red bold indicates >20% change)

| | Dice | | | | | | VSI | | | | | |
|---|---|---|---|---|---|---|---|---|---|---|---|---|
| Disease | Control | | MS | | ET | | Control | | MS | | ET | |
| Nuclei (volume) | FreeSurfer (n=5) | CNN (n=6) | FreeSurfer (n=14) | CNN (n=15) | FreeSurfer (n=11) | CNN (n=12) | FreeSurfer (n=5) | CNN (n=6) | FreeSurfer (n=14) | CNN (n=15) | FreeSurfer (n=11) | CNN (n=12) |
| Thalamus (5279) | 0.81 | 0.90* | 0.82 | 0.90* | 0.86 | 0.90* | 0.90 | 0.96 | 0.88 | 0.97* | 0.94 | 0.96 |
| Pul (1202) | 0.69 | **0.83*** | 0.67 | **0.83*** | 0.70 | **0.83*** | 0.93 | 0.94 | 0.84 | 0.93* | 0.88 | 0.93 |
| VLp (843) | 0.54 | **0.79*** | 0.57 | **0.78*** | 0.56 | **0.79*** | 0.96 | 0.94 | 0.93 | 0.93 | 0.92 | 0.93 |
| MD-Pf (670) | 0.56 | **0.83*** | 0.63 | **0.84*** | 0.68 | **0.83*** | 0.85 | 0.94 | 0.79 | 0.93* | 0.92 | 0.95 |
| VPI (337) | 0.45 | **0.64*** | 0.45 | **0.66*** | 0.45 | **0.65*** | 0.58 | **0.87*** | 0.57 | **0.88*** | 0.58 | **0.94*** |
| VA (286) | 0.40 | **0.71*** | 0.45 | **0.71*** | 0.48 | **0.67*** | 0.75 | 0.88 | 0.81 | 0.88* | 0.77 | 0.92* |
| AV (138) | 0.37 | **0.77*** | 0.38 | **0.71*** | 0.47 | **0.70*** | 0.94 | 0.91 | 0.86 | 0.85 | 0.89 | 0.90 |
| VLa (116) | 0.21 | **0.64*** | 0.20 | **0.61*** | 0.19 | **0.65*** | 0.28 | **0.89*** | 0.28 | **0.88*** | 0.31 | **0.90*** |
| CM (110) | 0.33 | **0.67*** | 0.42 | **0.69*** | 0.35 | **0.65*** | 0.59 | **0.87*** | 0.60 | **0.90*** | 0.49 | **0.88*** |
| LGN (108) | 0.08 | **0.64*** | 0.09 | **0.59*** | 0.20 | **0.53*** | 0.81 | 0.80 | 0.83 | 0.78 | 0.87 | 0.85 |
| MGN (70) | 0.04 | **0.67*** | 0.06 | **0.69*** | 0.07 | **0.65*** | 0.77 | 0.87 | 0.77 | 0.83 | 0.87 | 0.88 |
| MTT (49) | | 0.50 | | 0.58 | | 0.60 | | 0.85 | | 0.85 | | 0.89 |
| Hb (28) | | 0.67 | | 0.67 | | 0.69 | | 0.87 | | 0.84 | | 0.87 |

\* p < 0.05 FreeSurfer vs. CNN

Images from a representative MS patient acquired using two different contrasts (WMn-MPRAGE and CSFn-MPRAGE) along with their overlaid segmentations are shown in Figure 5. We can clearly see that the network segmented the thalamic nuclei fairly accurately in the CSFn-MPRAGE image, despite its seemingly poor intra-thalamic contrast. For the larger nuclei, Dice values for these two subjects ranged from 0.69-0.86 and 0.66-0.89 for WMn-MPRAGE and CSFn-MPRAGE, respectively, and for the smaller nuclei, Dice values ranged from 0.73-0.82 and 0.61-0.81, respectively. Note also that the presence of MS lesions (arrows) has not affected the network performance, attesting to the robustness of the training process which included healthy and MS patients.

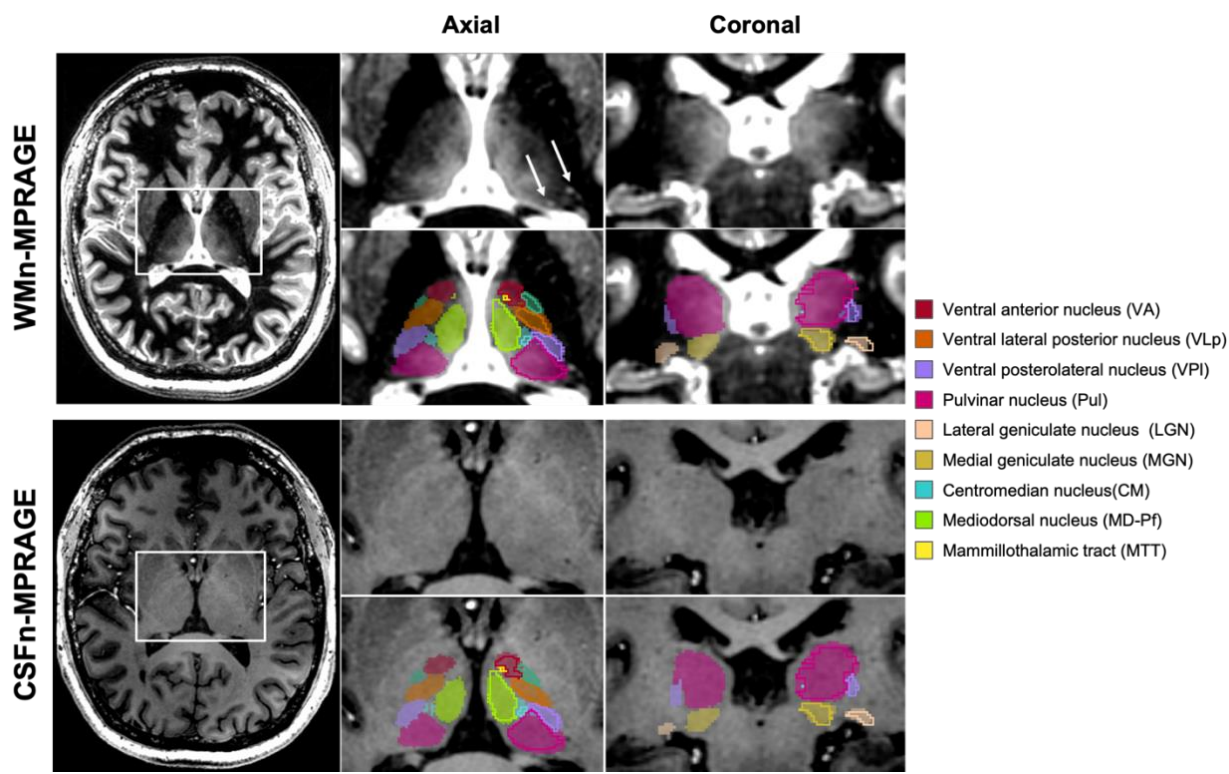

**Figure 5:** WMn-MPRAGE (top left) and CSFn-MPRAGE (bottom left) images from a patient with MS acquired at 7T. The zoomed inset images on the right show representative axial and coronal sections with and without automated segmentation overlays with manual segmentation shown for the left thalamus in outlines. Note the comparable thalamic nuclei segmentations for both cases despite the poor intra-thalamic nuclear contrast in the CSFn-MPRAGE image. MS lesions are shown as white arrows in the top inset image.

### 3.3 Network Robustness to Noise

Figure 6 shows the effect of noise addition on the performance of the proposed method. It can be seen that even though there is a gradual decline in Dice for whole thalamus and all nuclei (except for VPL and LGN), there is no significant reduction in accuracy until SNR = 11.4 (i.e. reduced ~ 2X from the baseline), showing that further acquisition speed-up (which usually comes at the cost of SNR) can be achieved. From SNR of 11.4 to SNR of 8 (~3X reduction from baseline), two nuclei (MTT and LGN) and the whole thalamus show significant decline whilst the remaining show a more modest decline.

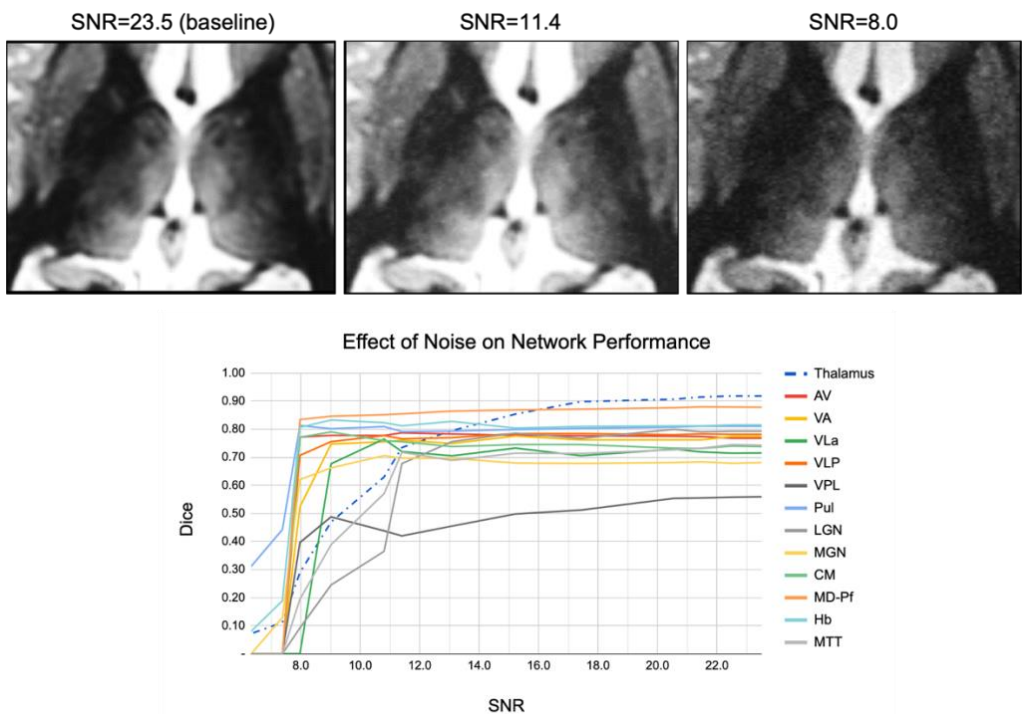

**Figure 6:** Effect of noise addition on the performance of the proposed method. Note the excellent performance on most nuclei for 2X lower SNR (SNR = 11.4) compared to baseline (SNR = 23).

### 3.4 Clinical Analysis in Multiple Sclerosis

Two-tailed t-test analyses showed no statistically significant difference between the automatically segmented nuclei volumes of left and right thalami except for MTT, and thus the average volume of left and right thalami was used for the analysis. An ANCOVA was conducted to compare absolute bilateral thalamic nuclear volumes between the two groups, controlling for age, gender, ICV, and PI. There was significant atrophy in MS patients in whole thalamus (F [1,22] = 9.5, $p$ = 0.005), AV (F [1,22] = 26.6, $p$ < 0.0001), Pul (F [1,22] = 18, $p$ = 0.0003), and MGN (F [1,22] = 17, $p$ = 0.0004). These were identical to the results observed using THOMAS segmented data except for an additional atrophy in Hb (F [1,22] = 9.1, $p$ = 0.006) which did not survive significance after Bonferroni correction.

## 4. DISCUSSION

Many studies have explored the use of CNNs in subcortical segmentation [43,46,50,51] but we believe this is the first work to use CNNs to segment the thalamic nuclei from structural MRI data. A single network for segmentation of WMn-MPRAGE data is introduced for 3T and 7T as well as for healthy controls and patients with MS and ET. Twelve thalamic nuclei including small structures such as the MTT, habenula, and lateral and medial geniculate nuclei were segmented in under a minute. A fine-tuning approach was employed to incorporate information learned from the WMn network to segment thalamic nuclei from CSFn-MPRAGE images, which have poor intra-thalamic contrast. While the Dice values were slightly lower than those from the corresponding WMn-MPRAGE images acquired on the same patients, they were significantly improved compared to FreeSurfer with significantly reduced processing times (1 minute vs. several hours). The proposed method also exhibited robustness to noise (up to 2X lower SNR compared to baseline).

Previous neuroimaging studies, albeit confined to the whole thalamus, have shown evidence of thalamic involvement in MS. Planche et al. [70] recently demonstrated atrophy of specific thalamic nuclei due to MS. Our method showed a statistically significant atrophy in patients with MS compared to healthy subjects for the whole thalamus as well as for AV, MGN, and pulvinar nuclei, comporting well with the results of Planche et al. [70]. The antero-ventral nucleus is a critical component in episodic memory and the circuit of Papez [70]. Our network successfully segmented this nucleus with a mean Dice of > 0.71 for both WMn-MPRAGE and CSFn-MPRAGE. The latter is of critical relevance for analyzing public databases such as the Alzheimer's disease neuroimaging initiative (ADNI) (which only has conventional CSFn-MPRAGE data), to study the effect of Alzheimer's disease on nuclei such as AV and MD, that are critically involved in episodic memory.

The recent shape-based segmentation method of Liu et al. [71] is one of the few methods that segment MPRAGE data using the Morel atlas convention. While their method generated 23 nuclei, the lateral and medial geniculate nuclei were notably absent in their output. During manual segmentation, we chose to merge some of the smaller nuclei with larger contiguous nuclei, resulting in a smaller number of reported nuclei compared to their method (e.g. MD is combined with Pf, and the pulvinar complex is segmented as a single nucleus in our manual segmentation). Our method's performance on the WMn-MPRAGE data shows comparable Dice values for most nuclei, lower Dice for VPL and VA nuclei and higher Dice for CM nucleus and MTT, compared to Liu's shape-based method. However, our network's performance on CSFn-MPRAGE data shows lower accuracy in comparison, presumably due to the lack of shape information, which could help in the face of poor intra-thalamic contrast. Our analysis had a much larger spread of cases encompassing healthy controls (n=13) as well as patients with ET (n=12) and MS (n=15), compared to 9 healthy subjects in Liu et al. The proposed method shows dramatically improved Dice accuracy over the Bayesian atlas-based method of Iglesias et al. [24] for all nuclei and disease types while significantly shortening processing times to under a minute from several hours.

The proposed CNN-based method has limitations that are common to most deep learning methods. Data diversity during the training phase is critical in creating a generalizable network. The current network was trained using images acquired from 3T and 7T and patients with MS and ET in addition to healthy subjects. However, since the network has not been exposed to images with metal artifacts from surgical clips or deep brain stimulation electrodes, it is likely to fail under those conditions and will require special training. The performance of this network on other diseases such as Alzheimer's disease needs to be evaluated. Limitations specific to our implementation include not taking advantage of 3D data; with sufficient 3D training data and memory, a 3D cascaded network will likely improve the accuracy of our method by fully leveraging the 3D structural information. Due to lack of manual segmentation data, the performance of the proposed method on 3T CSFn-MPRAGE data was not investigated. We also observed a slight reduction in performance in the CSFn-MPRAGE dataset compared to the WMn-MPRAGE dataset for the smaller nuclei. This could partly be due to the inherently lower intra-thalamic contrast in conventional CSFn-MPRAGE images. Future work will explore synthesizing WMn-MPRAGE images from CSFn-MPRAGE using contrast synthesis methods and then applying the WMn-MPRAGE optimized network for better accuracy.

**CONCLUSION**

We have proposed the use of a CNN-based cascaded multi-class multi-planar method for the segmentation of thalamic nuclei and evaluated it on images with different contrasts, and magnetic field strengths for both healthy and diseased populations. This method has been applied successfully to both advanced MR acquisition techniques with high intra-thalamic contrast (WMn-MPRAGE) and the more commonly used low intra-thalamic contrast sequences (CSFn-MPRAGE). Further, the effectiveness of this method in real-life applications has been investigated via a clinical analysis of volume atrophy in patients with MS.

**ACKNOWLEDGEMENTS**

We would like to acknowledge funding support from the National Institutes of Health (R21 AA023582-01) and the Arizona Alzheimer's Consortium.

ABBREVIATIONS
AV Anteroventral nucleus
CM Centromedian nucleus
DTI Diffusion tensor imaging
Hb Habenular nucleus
LGN Lateral geniculate nucleus
MD Mediodorsal nucleus
MGN Medial geniculate nucleus
MTT Mammillothalamic tract
Pul Pulvinar nucleus
VA Ventral anterior nucleus

VIM Ventralis intermedius nucleus

VLa Ventral lateral anterior nucleus

VLp Ventral lateral posterior nucleus

VPL Ventral posterolateral nucleus

WM White matter

WMn White matter nulled

CSF Cerebrospinal fluid

CSFn Cerebrospinal fluid nulled

MPRAGE Magnetization-prepared rapid gradient echo

**REFERENCES**

[1]  Aggleton JP, O'Mara SM, Vann SD, Wright NF, Tsanov M, Erichsen JT. Hippocampal-anterior thalamic pathways for memory: Uncovering a network of direct and indirect actions. Eur J Neurosci 2010;31:2292–307. https://doi.org/10.1111/j.1460-9568.2010.07251.x.

[2]  Steriade M, Llinás RR, Llinas RR. The functional states of the thalamus and the associated neuronal interplay. Physiol Rev 1988;68:649–742. https://doi.org/10.1152/physrev.1988.68.3.649.

[3]  Bodart O, Laureys S, Gosseries O. Coma and disorders of consciousness: Scientific advances and practical considerations for clinicians. Semin Neurol 2013;33:83–90. https://doi.org/10.1055/s-0033-1348965.

[4]  Stein T, Moritz C, Quigley M, Cordes D, Haughton V, Meyerand E. Functional connectivity in the thalamus and hippocampus studied with functional MR imaging. Am J Neuroradiol 2000;21:1397–401.

[5]  Chen Y, Shi B, Wang Z, Zhang P, Smith CD, Liu J. Hippocampus segmentation through multi-view ensemble ConvNets. Int. Symp. Biomed. Imaging, IEEE Computer Society; 2017, p. 192–6. https://doi.org/10.1109/ISBI.2017.7950499.

[6]  Parnaudeau S, Bolkan SS, Kellendonk C. The mediodorsal thalamus: An essential partner of the prefrontal cortex for cognition. Biol Psychiatry 2018;83:648–56. https://doi.org/10.1016/j.biopsych.2017.11.008.

[7]  Arts NJM, Walvoort SJW, Kessels RPC. Korsakoff's syndrome: A critical review.

Neuropsychiatr Dis Treat 2017;13:2875–90. https://doi.org/10.2147/NDT.S130078.

[8] Fama R, Rosenbloom MJ, Sassoon SA, Rohlfing T, Pfefferbaum A, Sullivan E V. Thalamic volume deficit contributes to procedural and explicit memory impairment in HIV infection with primary alcoholism comorbidity. Brain Imaging Behav 2014;8:611–20. https://doi.org/10.1007/s11682-013-9286-4.

[9] Benabid AL, Pollak P, Seigneuret E, Hoffmann D, Gay E, Perret J. Chronic VIM thalamic stimulation in Parkinson's disease, essential tremor and extra-pyramidal dyskinesias. Acta Neurochir Suppl (Wien) 1993. https://doi.org/10.1007/978-3-7091-9297-9_8.

[10] Minagar A, Barnett MH, Benedict RHBB, Pelletier D, Pirko I, Sahraian MA, et al. The thalamus and multiple sclerosis: modern views on pathologic, imaging, and clinical aspects. Neurology 2013;80:210–9. https://doi.org/10.1212/WNL.0b013e31827b910b.

[11] Braak H, Braak E. Alzheimer's disease affects limbic nuclei of the thalamus. Acta Neuropathol 1991;81:261–8. https://doi.org/10.1007/BF00305867.

[12] Lee J-Y, Jin E, Lee WW, Kim YK, Lee J-Y, Jeon B. Lateral geniculate atrophy in Parkinson's with visual hallucination: A trans-synaptic degeneration? Mov Disord 2016;31:547–54. https://doi.org/10.1002/mds.26533.

[13] Brodkey JA, Tasker RR, Hamani C, McAndrews MP, Dostrovsky JO, Lozano AM. Tremor cells in the human thalamus: Differences among neurological disorders. J Neurosurg 2004;101:43–7. https://doi.org/10.3171/jns.2004.101.1.0043.

[14] Benabid AL, Pollak P, Gao D, Hoffmann D, Limousin P, Gay E, et al. Chronic electrical stimulation of the ventralis intermedius nucleus of the thalamus as a treatment of movement disorders. J Neurosurg 1996;84:203–14. https://doi.org/10.3171/jns.1996.84.2.0203.

[15] Fama R, Sullivan E V. Thalamic structures and associated cognitive functions: Relations with age and aging. Neurosci Biobehav Rev 2015;54:29–37. https://doi.org/10.1016/j.neubiorev.2015.03.008.

[16] Elias WJ, Huss D, Voss T, Loomba J, Khaled M, Zadicario E, et al. A pilot study of focused ultrasound thalamotomy for essential tremor. N Engl J Med 2013;369:640–8. https://doi.org/10.1056/NEJMoa1300962.

[17] Krauth A, Blanc R, Poveda A, Jeanmonod D, Morel A, Székely G. A mean three-dimensional atlas of the human thalamus: Generation from multiple histological data. Neuroimage 2010;49:2053–62. https://doi.org/10.1016/j.neuroimage.2009.10.042.

[18] Jakab A, Blanc R, Berényi EL, Székely G. Generation of individualized thalamus target maps by using statistical shape models and thalamocortical tractography. Am J Neuroradiol 2012;33:2110–6. https://doi.org/10.3174/ajnr.A3140.

[19] Tourdias T, Saranathan M, Levesque IR, Su J, Rutt BK. Visualization of intra-thalamic nuclei with optimized white-matter-nulled MPRAGE at 7T. Neuroimage 2014;84:534–45. https://doi.org/10.1016/j.neuroimage.2013.08.069.

[20] Fischl B, Salat DH, Busa E, Albert M, Dieterich M, Haselgrove C, et al. Whole brain segmentation: Automated labeling of neuroanatomical structures in the human brain. Neuron 2002;33:341–55. https://doi.org/10.1016/S0896-6273(02)00569-X.

[21] FreeSurfer FB. B. Fischl. Neuroimage 2012;62:774–81.

[22] Patenaude B, Smith SM, Kennedy DN, Jenkinson M. A Bayesian model of shape and appearance for subcortical brain segmentation. Neuroimage 2011;56:907–22. https://doi.org/10.1016/j.neuroimage.2011.02.046.

[23] Heckemann RA, Hajnal J V., Aljabar P, Rueckert D, Hammers A. Automatic anatomical brain MRI segmentation combining label propagation and decision fusion. Neuroimage 2006;33:115–26. https://doi.org/10.1016/j.neuroimage.2006.05.061.

[24] Iglesias JE, Insausti R, Lerma-Usabiaga G, Bocchetta M, Van Leemput K, Greve DN, et al. A probabilistic atlas of the human thalamic nuclei combining ex vivo MRI and histology. Neuroimage 2018;183:314–26. https://doi.org/10.1016/j.neuroimage.2018.08.012.

[25] Smith SM, Jenkinson M, Woolrich MW, Beckmann CF, Behrens TEJJ, Johansen-Berg H, et al. Advances in functional and structural MR image analysis and implementation as FSL. Neuroimage 2004;23:S208–19. https://doi.org/10.1016/j.neuroimage.2004.07.051.

[26] Behrens TEJJ, Johansen-Berg H, Woolrich MW, Smith SM, Wheeler-Kingshott CAMM, Boulby PA, et al. Non-invasive mapping of connections between human

thalamus and cortex using diffusion imaging. Nat Neurosci 2003;6:750–7. https://doi.org/10.1038/nn1075.

[27] Wiegell MR, Tuch DS, Larsson HBWW, Wedeen VJ. Automatic segmentation of thalamic nuclei from diffusion tensor magnetic resonance imaging. Neuroimage 2003;19:391–401. https://doi.org/10.1016/S1053-8119(03)00044-2.

[28] Kumar V, Mang S, Grodd W. Direct diffusion-based parcellation of the human thalamus. Brain Struct Funct 2015;220:1619–35. https://doi.org/10.1007/s00429-014-0748-2.

[29] Mang SC, Busza A, Reiterer S, Grodd W, Klose AU. Thalamus segmentation based on the local diffusion direction: A group study. Magn Reson Med 2012;67:118–26. https://doi.org/10.1002/mrm.22996.

[30] Battistella G, Najdenovska E, Maeder P, Ghazaleh N, Daducci A, Thiran JP, et al. Robust thalamic nuclei segmentation method based on local diffusion magnetic resonance properties. Brain Struct Funct 2017;222:2203–16. https://doi.org/10.1007/s00429-016-1336-4.

[31] Su JH, Thomas FT, Kasoff WS, Tourdias T, Choi EY, Rutt BK, et al. Thalamus optimized multi atlas segmentation (THOMAS): Fast, fully automated segmentation of thalamic nuclei from structural MRI. Neuroimage 2019;194:272–82. https://doi.org/10.1016/j.neuroimage.2019.03.021.

[32] Haacke EM, Xu Y, Cheng YCN, Reichenbach JR. Susceptibility weighted imaging (SWI). Magn Reson Med 2004;52:612–8. https://doi.org/10.1002/mrm.20198.

[33] Abosch A, Yacoub E, Ugurbil K, Harel N. An assessment of current brain targets for deep brain stimulation surgery with susceptibility-weighted imaging at 7 tesla. Neurosurgery 2010;67:1745–56. https://doi.org/10.1227/NEU.0b013e3181f74105.

[34] Xiao YZ, Zitella LM, Duchin Y, Teplitzky BA, Kastl D, Adriany G, et al. Multimodal 7T imaging of thalamic nuclei for preclinical deep brain stimulation applications. Front Neurosci 2016;10:264. https://doi.org/10.3389/fnins.2016.00264.

[35] Stough J V., Ye C, Ying SH, Prince JL. Thalamic parcellation from multi-modal data using random forest learning. Int. Symp. Biomed. Imaging, 2013, p. 852–5. https://doi.org/10.1109/ISBI.2013.6556609.

[36] Glaister J, Carass A, NessAiver T, Stough J V., Saidha S, Calabresi PA, et al.

Thalamus segmentation using multi-modal feature classification: Validation and pilot study of an age-matched cohort. Neuroimage 2017;158:430–40. https://doi.org/10.1016/j.neuroimage.2017.06.047.

[37] Van Leemput K. Encoding probabilistic brain atlases using Bayesian inference. IEEE Trans Med Imaging 2009;28:822–37. https://doi.org/10.1109/TMI.2008.2010434.

[38] Iglesias JE, Augustinack JC, Nguyen K, Player CM, Player A, Wright M, et al. A computational atlas of the hippocampal formation using ex vivo, ultra-high resolution MRI: Application to adaptive segmentation of in vivo MRI. Neuroimage 2015;115:117–37. https://doi.org/10.1016/j.neuroimage.2015.04.042.

[39] Liu Y, D'Haese P-F, Newton AT, Dawant BM. Thalamic nuclei segmentation in clinical 3T T1-weighted Images using high-resolution 7T shape models. In: Webster RJ, Yaniv ZR, editors. Med. Imaging Image-Guided Proced. Robot. Interv. Model., vol. 9415, 2015, p. 94150E. https://doi.org/10.1117/12.2081660.

[40] Vassal F, Coste J, Derost P, Mendes V, Gabrillargues J, Nuti C, et al. Direct stereotactic targeting of the ventrointermediate nucleus of the thalamus based on anatomic 1.5T MRI mapping with a white matter attenuated inversion recovery (WAIR) sequence. Brain Stimul., vol. 5, 2012, p. 625–33. https://doi.org/10.1016/j.brs.2011.10.007.

[41] Sudhyadhom A, Haq IU, Foote KD, Okun MS, Bova FJ. A high resolution and high contrast MRI for differentiation of subcortical structures for DBS targeting: The fast gray matter acquisition T1 inversion recovery (FGATIR). Neuroimage 2009;47:T44--52. https://doi.org/10.1016/j.neuroimage.2009.04.018.

[42] Alvén J. Improving multi-atlas segmentation methods for medical images. Chalmers University of Technology, 2017.

[43] Brebisson A de, de Brebisson A, Montana G, De Brébisson A, Montana G. Deep neural networks for anatomical brain segmentation. IEEE Comput Soc Conf Comput Vis Pattern Recognit Work 2015;2015-Octob:20–8. https://doi.org/10.1109/CVPRW.2015.7301312.

[44] Jia H, Xia Y, Cai W, Fulham M, Feng DD. Prostate segmentation in MR images using ensemble deep convolutional neural networks. Int Symp Biomed Imaging

2017:762–5. https://doi.org/10.1109/ISBI.2017.7950630.

[45] Chen Y, Shi B, Wang Z, Sun T, Smith CD, Liu J. Accurate and consistent hippocampus segmentation through convolutional LSTM and view ensemble. Mach Learn Med Imaging 2017;10541 LNCS:88–96. https://doi.org/10.1007/978-3-319-67389-9_11.

[46] Moeskops P, Viergever MA, Mendrik AM, de Vries LS, Benders MJNLNL, Isgum I. Automatic segmentation of MR brain images with a convolutional neural network. IEEE Trans Med Imaging 2016;35:1252–61. https://doi.org/10.1109/TMI.2016.2548501.

[47] Guo Y, Liu Y, Oerlemans A, Lao S, Wu S, Lew MS. Deep learning for visual understanding: A review. Neurocomputing 2016;187:27–48. https://doi.org/10.1016/j.neucom.2015.09.116.

[48] Ram S, Majdi MS, Rodriguez JJ, Gao Y, Brooks HL. Classification of primary cilia in microscopy images using convolutional neural random forests. Southwest Symp. Image Anal. Interpret., IEEE Computer Society; 2018, p. 89–92. https://doi.org/10.1109/SSIAI.2018.8470320.

[49] Zhu H, Shi F, Wang L, Hung S-C, Chen M-H, Wang S, et al. Dilated dense U-Net for infant hippocampus subfield segmentation. Front Neuroinform 2019;13:30. https://doi.org/10.3389/fninf.2019.00030.

[50] Shaken M, Tsogkas S, Ferrante E, Lippe S, Kadoury S, Paragios N, et al. Sub-cortical brain structure segmentation using F-CNN's. Symp. Biomed. Imaging, IEEE Computer Society; 2016, p. 269–72. https://doi.org/10.1109/ISBI.2016.7493261.

[51] Milletari F, Ahmadi S-A, Kroll C, Plate A, Rozanski V, Maiostre J, et al. Hough-CNN: Deep learning for segmentation of deep brain regions in MRI and ultrasound. Comput Vis Image Underst 2016;164:92–102. https://doi.org/10.1016/j.cviu.2017.04.002.

[52] Kushibar K, Valverde S, González-Villà S, Bernal J, Cabezas M, Oliver A, et al. Automated sub-cortical brain structure segmentation combining spatial and deep convolutional features. Med Image Anal 2018;48:177–86. https://doi.org/10.1016/j.media.2018.06.006.

[53] Roth HR, Lu L, Seff A, Cherry KM, Hoffman J, Wang S, et al. A new 2.5D

representation for lymph node detection using random sets of deep convolutional neural network observations. Lect. Notes Comput. Sci., vol. 8673 LNCS, Springer Verlag; 2014, p. 520–7. https://doi.org/10.1007/978-3-319-10404-1_65.

[54] Prasoon A, Petersen K, Igel C, Lauze F, Dam E, Nielsen M. Deep feature learning for knee cartilage segmentation using a triplanar convolutional neural network. Lect. Notes Comput. Sci., vol. 8150 LNCS, 2013, p. 246–53. https://doi.org/10.1007/978-3-642-40763-5_31.

[55] Tajbakhsh N, Shin JY, Gurudu SR, Hurst RT, Kendall CB, Gotway MB, et al. Convolutional neural networks for medical image analysis: Full training or fine tuning? IEEE Trans Med Imaging 2016;35:1299–312. https://doi.org/10.1109/TMI.2016.2535302.

[56] Ronneberger O, Fischer P, Brox T. U-Net: Convolutional networks for biomedical image segmentation. MICCAI, vol. 9351, Springer; 2015, p. 234–41. https://doi.org/10.1007/978-3-319-24574-4_28.

[57] Ioffe S, Christian Szegedy. Batch normalization: Accelerating deep network training by reducing. J Mol Struct 2015;1134:63–6. https://doi.org/10.1016/j.molstruc.2016.12.061.

[58] Srivastava N, Hinton G, Krizhevsky A, Salakhutdinov R, Sutskever I, Salakhutdinov R. Dropout: A simple way to prevent neural networks from overfitting. J Mach Learn Res 2014;15:1929–58.

[59] He K, Zhang X, Ren S, Sun J. Deep residual learning for image recognition. Proc. IEEE Comput. Soc. Conf. Comput. Vis. Pattern Recognit., vol. 2016- Decem, IEEE Computer Society; 2016, p. 770–8. https://doi.org/10.1109/CVPR.2016.90.

[60] Maas AL, Hannun AY, Ng AY. Rectifier nonlinearities improve neural network acoustic models. Int. Conf. Mach. Learn., vol. 30, 2013, p. 3.

[61] Novikov AA, Lenis D, Major D, Hladuvka J, Wimmer M, Buhler K, et al. Fully convolutional architectures for multiclass segmentation in chest radiographs. IEEE Trans Med Imaging 2018;37:1865–76. https://doi.org/10.1109/TMI.2018.2806086.

[62] Crum WR, Camara O, Hill DLGG. Generalized overlap measures for evaluation and validation in medical image analysis. IEEE Trans Med Imaging 2006;25:1451–61. https://doi.org/10.1109/TMI.2006.880587.

[63] Drozdzal M, Vorontsov E, Chartrand G, Kadoury S, Pal C. The importance of skip connections in biomedical image segmentation. Lect Notes Comput Sci 2016;10008 LNCS:179–87. https://doi.org/10.1007/978-3-319-46976-8_19.

[64] Niemann K, Mennicken VR, Jeanmonod D, Morel A. The morel stereotactic atlas of the human thalamus: Atlas-to-MR registration of internally consistent canonical model. Neuroimage 2000;12:601–16. https://doi.org/10.1006/nimg.2000.0650.

[65] Chollet F. Keras: the Python deep learning API 2017. https://keras.io/ (accessed May 15, 2020).

[66] Kingma DP, Ba JL, Diederik K, Ba JL. Adam: A method for stochastic optimization. Int. Conf. Learn. Represent., vol. 1631, 2014, p. 58–62. https://doi.org/10.1063/1.4902458.

[67] Dice LR. Measures of the amount of ecologic association between species. Ecology 1945;26:297–302. https://doi.org/10.2307/1932409.

[68] Hoult DI, Lauterbur PC. The sensitivity of the zeugmatographic experiment involving human samples. J Magn Reson 1979;34:425–33. https://doi.org/10.1016/0022-2364(79)90019-2.

[69] Edelstein WA, Glover GH, Hardy CJ, Redington RW. The intrinsic signal-to-noise ratio in NMR imaging. Magn Reson Med 1986;3:604–18. https://doi.org/10.1002/mrm.1910030413.

[70] Planche V, Su JH, Mournet S, Saranathan M, Dousset V, Han M, et al. White-matter-nulled MPRAGE at 7T reveals thalamic lesions and atrophy of specific thalamic nuclei in multiple sclerosis. Mult Scler J 2019:135245851982829. https://doi.org/10.1177/1352458519828297.

[71] Liu Y, D'Haese PF, Newton AT, Dawant BM. Generation of human thalamus atlases from 7 T data and application to intrathalamic nuclei segmentation in clinical 3 T T1-weighted images. Magn Reson Imaging 2020;65:114–28. https://doi.org/10.1016/j.mri.2019.09.004.